%% file: main.tex
\documentclass[sigconf]{acmart}

\usepackage{color}
\usepackage{url}
\usepackage{balance}
\usepackage{float}
\usepackage{makecell}
\usepackage{graphicx} 
\usepackage{enumitem}
\usepackage{multirow}
\usepackage{array}
\usepackage{booktabs}
\usepackage{soul}

\makeatletter
    \newcommand{\linkdest}[1]{\Hy@raisedlink{\hypertarget{#1}{}}}
\makeatother
\begin{document}
\settopmatter{printacmref=false} 
\renewcommand\footnotetextcopyrightpermission[1]{} 

\pagestyle{empty}
\title{
Data Quality Awareness: A Journey from Traditional Data Management to Data Science Systems
}

\author{Sijie Dong$^1$, Soror Sahri$^1$, Themis Palpanas$^2$} 
\address{
$^1$Université Paris Cité, LIPADE, F-75006 Paris, France $\>\>$ \texttt{\{sijie.dong@etu.u-paris.fr, soror.sahri@parisdescartes.fr\}} \\
$^2$University of Paris Cité \& French University Institute (IUF), France $\>\>$ \texttt{themis@mi.parisdescartes.fr} \\
}

\begin{abstract}

Artificial intelligence (AI) has transformed various fields, significantly impacting our daily lives. A major factor in AI’s success is high-quality data. In this paper, we present a comprehensive review of the evolution of data quality (DQ) awareness from traditional data management systems to modern data-driven AI systems, which are integral to data science. 
We synthesize the existing literature, highlighting the quality challenges and techniques that have evolved from traditional data management to data science including big data and ML fields. 
As data science systems support a wide range of activities, our focus in this paper lies specifically in the analytics aspect driven
by machine learning. We use the cause-effect connection between the quality challenges of ML and those of big data to allow a more thorough understanding of emerging DQ challenges and the related quality awareness techniques in data science systems. 
To the best of our knowledge, our paper is the first to provide a review of DQ awareness spanning traditional and emergent data science systems. 
We hope that readers will find this journey through the evolution of data quality awareness insightful and valuable.

\end{abstract}







\maketitle

\input{Sections/introduction}
\input{Sections/NewBackground}
\input{Sections/NewDSdimensions}
\input{Sections/MLSystems}
\input{Sections/OpenChallenges}
\input{Sections/conclusion}
\vspace{0.3cm}
\noindent{\bf Acknowledgements}
We would like to thank Dr. Divesh Srivastava for his several comments that helped improve our paper.

\bibliographystyle{ACM-Reference-Format}
\bibliography{main}


\end{document}

%% file: Sections/introduction.tex
\section{Introduction}
\label{sec:intro}
The rapid emergence of data-centric technologies, particularly in the big data and ML field, has increased attention to the challenge of data quality, prompting a need for DQ awareness in emergent data science systems. 
Although recent research work still considers the importance of adapting existing data quality characteristics and frameworks from traditional data management practices, it is still somewhat unclear how to adapt them to improve DQ awareness in these systems. 
Data-driven AI systems, which are integral to data science, emphasize the integration of adaptive and scalable quality measures capable of dynamically responding to evolving data landscapes and user requirements. 
This is crucial for ensuring that data inconsistencies do not compromise the accuracy and reliability of AI outputs. Effective data quality management in these systems involves real-time integration of quality measures that adapt to changes in data and user needs.

Our journey from traditional data management to data science, including big data and machine learning (ML), reflects an evolving landscape of data quality challenges. 
The singular focus of traditional data management systems on meeting the needs of immediate users often oversimplified the complexity of data quality issues. 
In a big data context, data quality issues are multiplying due to (i) large datasets that grow to the range of Tera- and Peta-Bytes, (ii) different types of data that existing data quality dimensions and evaluation methods cannot cope with, and (iii) real-time and time-evolving data, which may alter the data characteristics and consequently the insight derived from it. 
In \cite{Saha2014DataQT}, Saha and Srivastava present two main areas where data quality management challenges arise in the big data environment: (i) discovering data quality semantics and data repairing; and (ii) the trade-off between accuracy and efficiency using various computing models. 
This emphasizes the big data quality dimensions are more complex which leads to new techniques to assess data quality such as using big data platforms. 

The ability to extract value from big data depends on data analytics which includes business intelligence, data visualization, machine learning, and statistical analysis. In this paper, we focus on machine learning which is considered the core of the big data revolution \cite{cappiello2018quality,polyzotis2017data}. This synergy between big data and machine learning allows data science to transform big data into insights, decisions, and predictions. However, the complex configurations of the datasets used in data science systems and the diverse backgrounds of ML practitioners introduce new data quality challenges. 

Despite the rich literature on data quality in both big data \cite{cappiello2018quality,cai2015challenges} and machine learning \cite{10.1145/3592616} fields, as well as on the challenges of ML in conjunction with big data \cite{7906512,polyzotis2017data}, there is a gap in the literature that specifically explores the interplay between data quality issues and the unique challenges that arise from linking ML and big data within the scope of data science systems. 
Motivated by the need for data quality awareness in emergent data science systems, our paper aims to take readers on an enlightening journey through the evolution of data quality awareness from traditional to data science systems. 
We particularly focus on ML pipelines as a key component of data science systems. 
We emphasize the essential baggage from big data and traditional data management, including various quality awareness techniques, for effectively navigating the emerging data quality challenges in ML pipelines. 
By connecting these techniques with those of machine learning, we gain a deeper understanding of the challenges at hand. 

The rest of the paper provides an overview of fundamental data quality concepts in Section 2. 
Section 3 presents the main DQ techniques in traditional systems. 
Section 4 explores the DQ issues in big data systems, including new dimensions and techniques. 
Section 5 discusses quality awareness in ML pipelines. 
Finally, Section 6 summarizes the key findings and suggests future research directions to advance data quality awareness in modern data science systems.


%% file: Sections/NewBackground.tex

\section{Foundational concepts of Data Quality}
\textit{Data quality} refers to the extent to which data is suitable for a specific task, emphasizing its actual utility and relevance to the context in which it is used. Data quality is often measured by its "fitness for use" in supporting operations, decision-making, and planning in various contexts \cite{redman1997data}. To evaluate and ensure data quality, two main ingredients are used: \textit{DQ Dimensions} and \textit{DQ Metrics}. \textit{DQ Dimensions} represent various aspects of data that determine its overall quality, including attributes such as accuracy, completeness, consistency, and timeliness. These dimensions provide a framework to assess how well data meets its intended purpose \cite{wang1996beyond, batini2006data}. \textit{DQ Metrics} are quantifiable measures that evaluate these dimensions by assigning scores or percentages, helping to objectively assess data quality. Metrics enable comparison between datasets or across time, aiding in the continuous monitoring and improvement of data quality \cite{pipino2002data, wang1996beyond}.


\textit{Data Quality Awareness} involves an organization’s recognition of essential quality dimensions and metrics as a foundation for high-quality data. However, awareness alone cannot achieve or sustain these standards. Batini et al. \cite{batini2009methodologies} describe a data quality methodology as a structured framework guiding the selection and application of tools and techniques tailored to specific needs, thus translating awareness into practical steps for maintaining data quality. Supported by such a methodology, DQ awareness extends beyond understanding metrics to foster a structured approach for implementing and sustaining standards. Continuous improvement and assessment efforts are then essential to enhance source quality and meet or exceed user expectations \cite{cai2015challenges, batini2009methodologies}.

\textit{Data Quality Assessment} is the process of systematically evaluating data against relevant quality dimensions to determine its adequacy for a given purpose. This involves identifying data anomalies, errors, and inconsistencies using predefined metrics and tests. Effective data quality assessment not only pinpoints defects but also elucidates their causes, enabling targeted interventions to bolster data quality~\cite{10.1145/1541880.1541883,song2020data}.

\textit{Data Quality Improvement} aims to elevate the standards of data quality to meet or exceed new targets through the implementation of corrective measures derived from assessment results. Improvement strategies are typically categorized into two main approaches: data-driven and process-driven. Data-driven techniques directly update data values, like data cleansing. Process-driven techniques focus on redesigning data creation or modification processes, such as adding data format validation steps.

\section{Data Quality Awareness in Traditional Data Management}
Traditional data management systems incorporate various techniques to provide data quality awareness. Literature shows that most of these techniques are designed from the perspective of data producers or data sources, focusing on modeling and measuring the quality of data at its origin. 
However, to effectively address data quality challenges, it is crucial to also consider the requirements of data consumers as well, incorporating techniques for modeling user requirements on data quality and ensuring that the data meets these expectations. 
Following, we emphasize the most commonly used techniques in our categorization, exploring them from both data and user perspectives.

\subsection{Quality Awareness from Data Perspective } 
The data perspective for quality awareness focuses on data characterization. \textit{Dimensions} are used to characterize various data properties \cite{sidi2012data}. 
Literature categorizes the most common DQ dimensions into intrinsic, accessibility, contextual, and representational quality \cite{DBLP:journals/jmis/WangS96}. 
The intrinsic dimensions are context-free and related to internal properties of data (e.g., accuracy); the accessibility dimensions are related to data access (e.g., availability); the representational dimensions are related to the design of the data (e.g., consistency); and the contextual ones, as their names indicate, are context-dependent (e.g., timeliness). 
In~\cite{jayawardene2013curse, jayawardene2015analysis}, quality dimensions were divided into two types: 
(i) those with a declarative perspective to explain services (e.g., accuracy, completeness, and timeliness); 
(ii) and those with a perceptual perspective, which is convenient for perceptual judgment of service usage (consistency, accessibility, and responsiveness). 
To understand data characteristics and measure their data quality, data profiling techniques are mostly used, at various levels including: (i) attribute/tuple level such as missing value and domain violation; 
(ii) single relation such as violation of business rules; (iii) multiple relations as referential integrity violation; and (iv) multiple sources as inconsistent duplicate tuples. 
Existing data quality profiling techniques include defining new data quality rules (e.g., Functional Dependencies) and identifying data quality problems (e.g., inconsistent data)~\cite{DBLP:journals/sigmod/Naumann13, oliveira2006data, abedjan2017data, chiang2008discovering, ilyas2015trends}. 

The concept of data quality profiling in~\cite{DBLP:conf/iq/Berti-Equille04} is achieved by associating quality contracts with data sources. A set of contracts related to one or more data sources forms a quality profile. These profiles are used to negotiate quality requirements with the data source's wrapper, ensuring that the query processing framework selects sources based on their quality characteristics.

Conditional DQ profiling, as proposed in \cite{YEGANEH201424}, associates the quality of specific data attributes with conditions specified in user queries.
For example, the quality of the \textit{Price} attribute is evaluated only when the \textit{Brand} is specified as \textit{Sony} in a query, as presented in \cite{YEGANEH201424}. This method allows for targeted assessments of data quality based on user-defined relevance, rather than a blanket evaluation of all attributes. In addition, when quality metadata is unavailable, broader quality metadata may be used to estimate attribute completeness. For instance, if the \textit{Price} attribute lacks completeness data, the overall dataset completeness can serve as a surrogate. However, this method can introduce significant errors; for example, a dataset might be 50\% complete overall while the Price attribute could be only 25\% complete, leading to discrepancies in data quality assessments.


\subsection{Quality Awareness from User perspective}
Quality-aware query processing techniques are largely used to ensure that the quality of data meets user requirements and preferences, this mainly involves \textit{query language extensions} and \textit{adapting query processing}.
\subsubsection{Query language extensions} They enable the integration of data quality considerations into query processing. 
They allow the expression of quality metrics and constraints across various quality dimensions in a simple and declarative manner.
In \cite{YEGANEH201424}, an SQL extension was proposed to model user preferences related to data quality through hierarchical prioritization. Additionally, a framework referred to as DQAQS was introduced to enhance user satisfaction by considering these preferences in query results. 
In \cite{10.1145/1577840.1577846}, an extension to XQuery incorporates domain-specific quality constraints, referred to as quality views (QV), during query processing. This work extends previous works of the same author that addressed completeness \cite{sampaio2007incorporating} and timeliness dimensions \cite{Dong2006ExpressingAP}. 

In~\cite{DBLP:conf/iq/Berti-Equille04}, the quality-extended query language XQual, based on a negotiation strategy, was introduced for selecting dynamic sources. XQual extends SQL with a Qwith operator to specify quality constraints through quality contracts and profiles, adapting query processing to meet defined quality dimensions (e.g., \textit{dataAge} and \textit{lastUpdate}). This approach was later applied to online skyline queries, incorporating graph methods to optimize quality-based results using the nearest neighbor search~\cite{DBLP:journals/debu/Berti-Equille06}. Recently, XQual was used to enforce data quality thresholds for training data in machine learning, ensuring that only data meeting specified quality standards is used~\cite{DBLP:conf/edbt/ComignaniNB20}.

\subsubsection{Adaptive query processing} It allows to dynamically handle data quality issues and provide more reliable query results. It involves creating multiple query execution plans that can be switched dynamically during runtime based on the quality of the data. In~\cite{Naumann:1999:QIH:645925.671356, DBLP:books/sp/Naumann02}, a distributed query planning (DQP) algorithm discards low-quality sources, ordering plans by completeness. 
System P~\cite{roth2007completeness} uses a completeness-driven approach where peers rank local plans by potential result size and prune based on user-set budget thresholds, balancing completeness and cost. Similarly,~\cite{YEGANEH201424} incorporates planning and optimization to assess each plan’s utility based on expected data quality, ensuring efficient handling of diverse data sources.





%% file: Sections/NewDSdimensions.tex
\section{Data Quality Awareness in Big Data} 
As data-centric technologies advanced, particularly with the advent of big data, the scope of data quality expanded. 
Following, we present the quality dimensions related to new data quality challenges and the impacts of big data characteristics on the big data dimensions.

\subsection{Big Data Quality Dimensions} 
Traditional quality dimensions (e.g., accuracy, completeness, consistency, etc.) are not sufficient to assess big data. 
In the context of big data, the meaning and calculation methods of these traditional dimensions undergo significant changes.

This section presents the common quality dimensions relevant to big data. We classify them according to the existing literature~\cite{DBLP:journals/dase/FirmaniMSB16,cappiello2018quality,cai2015challenges,10.1145/3603707}, into source-specific and users' perspective dimensions. 

\subsubsection{Source' Perspective dimensions} 
The UNEC (United Nations Economic Commission for Europe) classification identified three main types of data sources from the most structured to the least structured ones: process-mediated, machine-generated, and human-sourced \cite{DBLP:journals/dase/FirmaniMSB16}. Big data quality dimensions, characterized as \textit{source' perspective}, are then categorized according to these source types as follows:

\begin{itemize}[leftmargin=*]
    \item  \textit{Process-mediated} data sources, usually correspond to relational databases that provide structured data (e.g., business data with customer records). The main quality issues concern the values provided for records' attributes, e.g., incorrect values, duplicates, incomplete values, etc; and the related quality dimensions to assess the data are then: consistency, accuracy, freshness, etc. 
    \item \textit{Machine-generated} information sources, use sensors and machines to measure and record the events and situations in the physical world and produce machine-generated data that is often well-structured. Still, its size and speed can become increasingly important. The main quality issues consider the environment of such measures and records (e.g., the noise problems of machines, the environmental effects, etc). The related quality dimensions are mainly: accuracy, completeness, consistency, trustworthiness, and freshness. 
    \item \textit{Human-sourced} information sources correspond particularly to social networks that store human experiences, photographs, audio, videos, etc. In addition to some of the quality dimensions mentioned above, the ambiguity related to short text understanding can also be considered.
\end{itemize}

\subsubsection{Users' perspective dimensions} 
To better understand big data applications, quality dimensions are defined from the users' perspective. 
Two main data quality assessment approaches are considered, effective and context-dependent, to define the corresponding DQ dimensions.
The effective data quality assessment, emphasizes evaluating quality dimensions relevant to user interactions with data. Key quality dimensions include reliability, availability, usability, relevance, and presentation quality, as noted in \cite{cai2015challenges}. These dimensions reflect users' ease of accessing data, its usefulness, trustworthiness, alignment with users' expectations, and the overall improvement of their satisfaction. Structured hierarchically, they consist of sub-elements and indicators that refine their meaning. 
The context-dependent quality assessment, highlighted in~\cite{DBLP:journals/fgcs/ArdagnaCSV18}, emphasizes the adaptation of quality dimensions based on the specific context of data assessment, encompassing levels such as data source, data type, and intended application. For instance, accuracy in batch data differs from that in sensor streams. In structured datasets, completeness may refer to non-null values, while for image data, it could mean image clarity. In \cite{DBLP:journals/fgcs/MerinoCRSP16}, the adaptive quality model is extended by considering user requirements related to execution time and performance constraints, introducing the confidence dimension to address potential accuracy issues, and measuring data trustworthiness.
The importance of context is further illustrated by specific domain applications~\cite{fadlallah2023context, serra2022use}. 
Social media platforms prioritize timeliness and accuracy for sentiment analysis, while online news platforms value credibility to ensure that reported information is trustworthy~\cite{el2019big,el2019impact,tarmizi2019online}. 
In healthcare, accuracy, completeness, and privacy are critical for ensuring the reliability of clinical data and the success of predictive analytics used in disease forecasting~\cite{leitheiser2001data,liu2023review}. 
Similarly, in financial data analysis, accuracy and timeliness are crucial for robust market assessments and detecting fraud in real time~\cite{batini2009methodologies,tomar2023role}.

\subsection{Impact of Big Data Characteristics on Big Data Quality}
The works presented above, and much of the previous work in the area characterize big data quality by (i) investigating the relation between big data characteristics and data quality dimensions, 
(ii) identifying specific data quality dimensions, 
and (iii) analyzing how to evaluate them to constitute the basis of a big data quality assessment. However, there is still a gap between big data characteristics and data quality dimensions, and their impact on better understanding big data applications. 
This is due to the impact of data quality on the insight derived from it, which makes the data quality and its value conceptually different, despite their correlation \cite{DBLP:conf/webdb/AbiteboulDESWSS15}. 
Furthermore, data quality affects the whole big data pipeline for an application, including data acquisition, analysis, and different interpretations \cite{DBLP:journals/sigmod/SadiqDDFILMNZS17}. 
Hence, assessing the overall quality of big data for a given application (e.g. analytic task) remains a grand challenge. 

To fill the gap between Big Data Characteristics (BDC) and quality dimensions, previous work investigated the correlation between big data and data quality specifically for financial service organizations. \cite{Wahyudi2018RelatingBD} suggested that, in the financial domain where multiple data sources are used, data variety is the most important BDC that affects most DQ dimensions, e.g. accuracy, consistency, security timeliness, and completeness. It was also suggested that velocity is often correlated to timeliness due to the correspondence between the rapidity of generating and processing data and the timely use of data. 

\cite{ghasemaghaei2020role} studied the effect of data volume on DQ dimensions which in turn influence the overall effectiveness and adoption of big data analytics in business contexts. The DQ dimensions considered in this study are less common and are data diagnosticity, accessibility, security, and task complexity. Data diagnosticity corresponds to the valuable insight from data; data accessibility measures the easiness of data availability; data security is related to security issues when aggregating and analyzing big data; and task complexity is related to data processing tasks. Based on the theory of valence, used in economics and psychology to explain the relationships between individuals' behavior and the perceptions of risks and benefits, the authors studied the positive and negative roles of DQ dimensions to develop guidelines for business data practices. Their findings show that BDCs and in particular the data volume improve data diagnosticity and accessibility, positively impacting big data analysis. However, BDCs also increase data security concerns and task complexity, negatively affecting analysis. Indeed, the large volume of data raises security issues, and using frameworks like Hadoop for data aggregation in distributed environments poses additional risks~\cite{DBLP:books/sp/18/BertinoF18}.

The findings indicate that contextual data quality (DQ) dimensions, particularly timeliness and accessibility, are most closely correlated with Big Data Characteristics (BDCs) for user applications. However, there are several limitations in existing studies. For instance, DQ dimensions beyond data security and accessibility can also affect big data analytics usage. While BDCs often focus on data volume, the impacts of other dimensions like variety and velocity on data security and accessibility, and their influence on analytics usage, remain unclear. Additionally, most research is limited to the financial sector, lacking broader applications in other domains and types of big data applications. This narrow focus hinders generalization, as user validation is insufficient to broadly apply findings across various big data contexts.

\begin{figure}[tb]
   \centering
    \includegraphics[width=1\columnwidth]{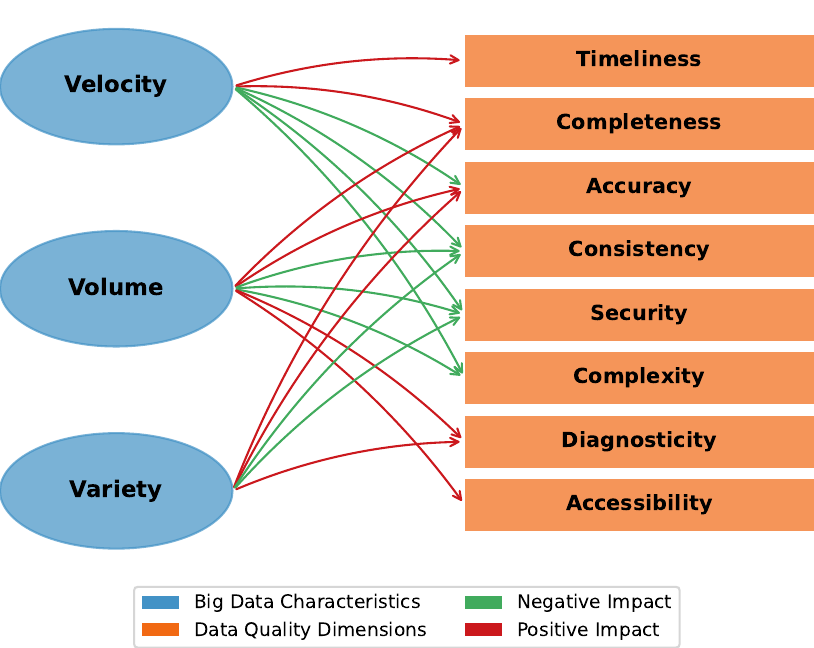}
    \vspace{-1.2\baselineskip}
   \caption{ The impact of the big data characteristics on data quality dimensions}
   \label{fig:impact_1}
\end{figure}

Even if some limitations can be pointed out in the above studies, their findings are useful for understanding the relationship between BDCs and DQ dimensions throughout user applications. Based on the above studies, we summarize the BDCs and their impact on DQ dimensions in Figure~\ref{fig:impact_1}.

\subsection{Quality Awareness Techniques for Big Data}
In this section, we present quality awareness techniques in a big data context, according to the V's characteristics, focusing on sampling, parallel processing, and incremental techniques.



\subsubsection{Parallel Computing}
Distributed frameworks such as Apache Hadoop and Apache Spark help manage data quality across large datasets by parallelizing the profiling and assessment tasks.
We showed in the section above, how parallel computing can resolve the problem that samples are not always representative of the whole large dataset, which is important to compute data quality measures. 
Parallel processing frameworks for big data can also be used for quality assessment that is not based on sampling. 
Several studies assess the performance and scalability of data quality (DQ) assessment tasks by implementing their solutions on big data computing frameworks. The main contributions indicate that using Spark can significantly enhance computational efficiency for very large data volumes~\cite{cisneros2021experimenting, schelter2018automating, cappiello2018quality}. For instance, Cisneros et al.~\cite{cisneros2021experimenting} investigate the performance of combining Spark and Python Pandas for data quality calculations compared to using them separately on different data sizes.

\subsubsection{Sampling and sketch techniques} 
\textit{Sampling techniques} help reduce the time required to calculate data quality by enabling the approximation of results. Commonly used methods include simple random sampling, systematic sampling, stratified sampling, cluster sampling, and reservoir sampling~\cite{liu2019approximate, 10.1214/aos/1176346500, 10665-264739, shomorony2014sampling, Vitter85randomsampling}. These techniques help determine sample sizes and select samples to effectively evaluate quality metadata considering the data source's features and the relevant quality dimensions. For instance, in~\cite{cappiello2018quality}, sampling improves calculation accuracy within the constraints of time or optimizes calculation time when accuracy meets user requirements. \cite{liu2019approximate} discusses the use of simple random and systematic sampling to assess data quality dimensions like completeness, accuracy, and timeliness, demonstrating that systematic sampling is preferable for accuracy and completeness, while simple random sampling suits timeliness assessment best. Further, \cite{taleb2016big} explores bootstrap sampling as part of a Big Data Quality (BDQ) evaluation scheme, which uses sampled data to profile the dataset and choose suitable quality metrics, particularly for accuracy, completeness, and consistency. 
The subsequent work in~\cite{taleb2018big} applies the Bag of Little Bootstrap (BLB) technique to enhance efficiency in processing unstructured data.

Sampling heterogeneous data requires preparation to ensure it is suitable for quality assessment. 
Taleb et al.~\cite{taleb2018big} proposed methods for preparing unstructured data like text, images, and videos for quality assessment. For instance, text data undergoes text mining, while video data requires feature extraction. This preparation identifies useful information that informs subsequent quality assessments by enabling the selection of key features. For remote sensing images, Wang et al.~\cite{wang2020multi} introduced a multi-level non-uniform spatial sampling method that improves accuracy assessment by considering spatial auto-correlation and heterogeneity. Additionally, other methods like multi-layer spatial sampling and BLB sampling are used, while random sampling is effective for faster evaluations with smaller datasets.

Overall, sampling is integrated into big data frameworks to accommodate large datasets, often necessitating innovative approaches like block-based sampling methods that utilize MapReduce to manage data scale and distribution effectively~\cite{he2017sampling}.

\textit{Sketch Techniques}, compared to sampling techniques, also greatly reduce the size of an input dataset. 
The difference is that sketching generates an approximate, compact data summary, which retains properties of interest. 
In contrast, sampling does not guarantee the preservation of such properties when selecting a subset of the data~\cite{DBLP:journals/cacm/Cormode17}. Sketch techniques are considered fast, easy to parallelize, and can provide high approximation accuracy~\cite{Cormode2010SketchTF, Dobra2004SketchBasedMP}. They are often used to efficiently find approximate answers to some online analytical processing queries, supported on big data (e.g., Pig, Hive, Spark SQL), with some reasonable guarantees on the quality. 
This is the case for many applications with online requirements that want to get results fast, as data-stream processing applications~\cite{Li2018ApproximateQP}.

\subsubsection{Data fusion and integration Techniques} These techniques involve combining data from various sources to create a coherent dataset. In the big data environment, data can originate from multiple sources such as social media, internal enterprise databases, and IoT devices, and these data can vary in structure, format, and quality \cite{khan2014big}. The goal of data fusion and integration is to resolve inconsistencies between these datasets, such as mismatches in timestamps, differences in data formats, and potential data duplication or conflicts~\cite{dong2013big}.
A large body of research has focused on data fusion and integration techniques, using methods such as data mapping, normalization, and transformation algorithms to ensure accuracy and consistency \cite{bleiholder2009data}. Metadata is often used to align the structure and semantics of different sources \cite{halevy2006data}, and quality assessments are often integrated to meet standards like integrity and accuracy. Readers can refer to existing studies for more in-depth insights~\cite{dong2013big,dong2009integrating,esmaelizadeh2024integrating}. 

\subsubsection{Incremental Big Data Quality Assessment.} 
The pace at which data is growing makes some data outdated, and consequently, the DQ assessment including profiling, very challenging. 
Indeed, data profiling methods should efficiently process such data growing, and without profiling the whole dataset again. 
Moreover, data quality metrics should be updated continuously. 
To improve the computation of the updated data, \cite{abedjan2017data} suggested performing incremental and continuous profiling. Incremental profiling updates data based on periodic changes from previous results, while continuous profiling updates data on the fly as data is entered. DQ metrics can be updated in three ways: 
on-demand, periodically or event-driven~\cite{cappiello2018quality, schelter2019differential}.
In addition, interactive profiling, including online profiling~\cite{DBLP:journals/sigmod/Naumann13}, improves user satisfaction by considering user quality requirements and displaying intermediate results from the interaction between users and applications. This allows users to make decisions based on early profiling results. Additionally, ~\cite{zernichow2017usability} proposed a visual data profiling interface to enable user interaction for data cleaning, error detection, and transformations.

%% file: Sections/MLSystems.tex
\section{Quality Awareness in ML Pipelines}
Data analytics, a cornerstone of big data, relies heavily on the quality of data to deliver accurate and actionable insights. Machine learning, as a core approach within data analytics, leverages these insights to develop predictive models and automate complex tasks, making it integral to modern data science systems. 
These emergent systems demand an iterative and integrated approach to data quality, addressing issues such as label quality, data drift, data imbalance, and incorrect data entries throughout the entire ML pipeline. 

\subsection{ML-based quality dimensions}
DQ dimensions tailored to ML pipelines cover the entire lifecycle, from data preparation to model training and monitoring. In this section, we review the literature on data quality dimensions in machine learning (ML) pipelines, based on various factors~\cite{polyzotis2018data, neutatz2021cleaning, daloisio2022modeling, mohammed2024data}. While these studies offer valuable classifications of data quality dimensions, their frameworks lack to emphasize the cause-effect connection between
the quality challenges of ML and those of big data to allow a more
thorough understanding of emerging data quality challenges in data science systems and particularly in ML pipelines. To address this, we synthesized the existing techniques in Section \ref{MLtechniques}, based on the following five main categories that consolidate and expand upon previous quality frameworks, in particular those in~\cite{mohammed2024data,d2022modeling,zhou2024surveydataqualitydimensions, 10.1145/3592616}. Table \ref{tab:quality_dimensions} presents these categories alongside their related quality-aware techniques.
~

\subsubsection{Data-based Dimensions}
The data considered here is used as input for each ML component. 
\cite{studer2021crispmlq} distinguishes between training data and serving data.
The training data refers to the dataset used during the development phase to train an ML component, while the serving data refers to the dataset used during the deployment phase to make predictions in real time.


Traditional data quality dimensions~\cite{jayawardene2015analysis}, such as representativeness, correctness, completeness, currentness, and intra-consistency, are essential to ensuring that the data accurately reflects real-world scenarios, is free from errors, and is both comprehensive and up-to-date.  
In addition to these conventional dimensions, machine learning introduces several unique data quality dimensions that are critical to ensuring the integrity and fairness of the model's performance. 
For example, \textit{Train/Test Independence} is a key consideration, where it is imperative that the training and testing datasets remain independent of each other. 
This prevents data leakage, which could otherwise artificially inflate performance metrics and lead to misleading conclusions about the model's effectiveness.
Another critical dimension is \textit{Balancedness}, which refers to the need for a balanced distribution of classes or categories within the dataset. An imbalanced dataset, where certain classes are overrepresented, can result in a biased model that performs poorly on underrepresented classes. Ensuring balanced data helps maintain fairness and accuracy across all predictions.
Lastly, the \textit{Absence of Bias} addresses the need to ensure that the dataset does not contain any inherent biases that could lead to unfair or unethical outcomes. 

\subsubsection{Model-based Dimensions}
The quality of an ML model is influenced by several factors, including the specific task being addressed (e.g., classification, clustering, regression, anomaly detection, dimensionality reduction), the type of model (neural network, decision tree, etc.), the data used for building the model (i.e., training), and evaluating the developed artifacts, as well as the manner in which the data is separated for training and validation \cite{Wagner_2015, Siebert_2020}. 

 Dimensions like performance, robustness, scalability, model complexity, and resource demand~\cite{jayawardene2015analysis} help ensure that the model operates effectively across various scenarios and environments. In addition to these conventional metrics, \textit{Fairness} and \textit{Explainability} are increasingly important for building models that are not only technically sound but also ethically responsible and trustworthy. \textit{Fairness} ensures that the model's decisions are unbiased and do not systematically disadvantage any particular group or individual. Addressing fairness allows for preventing discriminatory outcomes and promoting ethical AI use. \textit{Explainability} refers to the model’s ability to provide clear and understandable reasons for its decisions, which is important for gaining user trust and ensuring transparency. 
 
\subsubsection{Process-based Dimensions} 
The quality of machine learning systems depends not only on the data and models but also on the robustness of the processes used to develop, deploy, and maintain these systems. Effective process management ensures that the system remains reliable, efficient, and secure throughout its lifecycle. Key dimensions in this category, as highlighted by \cite{mohammed2024data} under the system facet,
include recoverability, portability, efficiency, transparency, traceability, cost, accessibility, ease of manipulation, and security. For example, traceability ensures that every step in the system's lifecycle can be audited, making it easier to identify and resolve issues, while efficiency focuses on optimizing resource usage such as computational power and time, ensuring the system runs smoothly and effectively. These dimensions are vital for ensuring the robustness, adaptability, and security of machine learning systems, particularly as they scale or evolve.



\subsubsection{Use case and context-based Dimensions}
Modeling quality requirements in ML pipelines based on specific use cases and application contexts is presented in \cite{Wagner_2015, Siebert_2020} to emphasize the importance of tailoring quality dimensions to the particular needs of ML applications. For instance, healthcare applications require stringent data privacy measures and high model accuracy. However, if the data used in these applications lacks contextual relevance, such as using a generic dataset for specialized medical diagnoses, this can lead to significant performance issues and may compromise patient safety. Addressing these issues helps ensure that ML models are effectively aligned with their intended applications, taking into account dimensions such as \textit{value}, \textit{contextual relevance}, and \textit{use case specificity}. Such alignment leverages trust in ML applications, highlighting the importance of continuous assessment and adaptation of quality requirements throughout the ML pipeline, as presented in \cite{10.1145/3592616}. Understanding these dimensions helps to identify various factors that can impact the quality of ML components (e.g., classification). Indeed, based on existing use cases, the various categories of quality dimensions identified can ensure that the ML pipeline effectively meets its intended objectives \cite{Siebert_2020}.

\subsubsection{Stakeholders-based Dimensions}
Data quality challenges related to stakeholders in ML pipelines are mainly presented in \cite{10.1145/3592616, polyzotis2017data}. The important role of ML practitioners, including data engineers, data scientists, and domain experts, should be emphasized to ensure both effectiveness and ethical responsibility in ML pipelines. This allows the data to align with specific use case requirements, mitigating the risk of using data in ways that may be ethically misaligned with the original intent of data curators. \textit{Ethical alignment} is then a key quality dimension allowing data usage to adhere to the ethical standards set by the data curators and comply with legal and social norms. For instance, in contexts where data can perpetuate biases, all stakeholders must ensure that their work supports fairness and accountability. In addition, all stakeholders from data curators to end-users, must have a clear understanding of how data is processed and how model decisions are made. For this aim, \textit{transparency} is another key dimension to build trust and facilitate collaboration across diverse users including developers, business analysts, and policymakers.

\subsection{Quality Awareness Techniques for ML Pipelines}
\label{MLtechniques}

Following, we present the ML quality techniques and emphasize their connection to big data challenges.

\begin{table*}[tb] 
\renewcommand{\arraystretch}{1.1} 
\setlength{\tabcolsep}{8pt} 
\centering 
\caption{Classification of Quality Dimensions and Techniques Across ML Pipeline Stages}  
\label{tab:quality_dimensions} 
\scriptsize 
\begin{tabular}{c|c|c|c|c|c|c} 
\hline 
\multirow{2}{*}{\textbf{ML Stages}} & \multirow{2}{*}{\textbf{Techniques}} & \multicolumn{5}{c}{\textbf{Quality Dimensions in the ML Pipeline}} \\ 
\cline{3-7} 
& & \textbf{\makecell{Data-based\\Dimensions}} & \textbf{\makecell{Model-based\\Dimensions}} & \textbf{\makecell{Process-based\\Dimensions}} & \textbf{\makecell{Use Case/\\Context-based\\Dimensions}} & \textbf{\makecell{Stakeholder-\\based\\Dimensions}} \\ 
\hline  

\multirow{5}{*}{\textbf{\makecell{Data \\Preparation}}}  
& \textit{Sampling} & \makecell{Representativeness,\\Balancedness} & & & Contextual Relevance & \\ 
\cline{2-7} 
& \textit{Data Validation} & Correctness, Completeness & & & & \\ 
\cline{2-7} 
& \textit{Data Cleaning} & Correctness, Completeness & & & & \\ 
\cline{2-7} 
& \textit{Data Imputation} & \makecell{Correctness, Completeness,\\Intra-Consistency} & & & & \\ 
\cline{2-7} 
& \textit{Data Labeling} & \makecell{Correctness, Absence of Bias} & & & & \\ 
\hline  

\multirow{3}{*}{\textbf{\makecell{Model\\ Training}}}      
& \textit{Feature Engineering} & \makecell{Representativeness, Train/Test\\Independence, Balancedness} & \makecell{Performance,\\Model Complexity} & & Use Case Specificity & \\ 
\cline{2-7} 
& \textit{Fairness and Bias Checking} & Absence of Bias & Fairness & & & Ethical Alignment \\ 
\cline{2-7} 
& \textit{Data Augmentation} & Balancedness & Fairness & & Use Case Specificity & \\ 
\hline  

\multirow{2}{*}{\textbf{\makecell{Model \\Validation \&\\ Monitoring}}}  
& \textit{Data Drift Detection} & Currentness & Robustness & Real-time Performance & & \\ 
\cline{2-7}  
& \textit{Monitoring Model} & & \makecell{Performance, Scalability,\\Trust} & \makecell{Reliability, Documentation\\Quality, Auditability,\\Reproducibility} & Trust & Transparency \\ 
\hline  
\end{tabular} 
\end{table*}

\subsubsection{Data Validation and Cleaning}

\textit{Data Validation} techniques verify that data conforms to certain standards~\cite{gupta2021data} and detect anomalies before they impact model performance. 
One fundamental technique is the use of \textit{descriptive statistics}, which involves calculating metrics such as mean, median, variance, and standard deviation to summarize the central tendency, dispersion, and shape of the data distribution~\cite{huber2011robust}. 
Recent advancements in this area include robust statistical methods that improve the detection of outliers and anomalies in large-scale datasets~\cite{she2011outlier}.
\textit{Schema validation} ensures that the data adheres to a predefined schema, checking for consistency in data types, ranges, and formats~\cite{abiteboul1995foundations}. Recent schema evolution techniques automatically adapt schemas to handle changes in data formats over time~\cite{baylor2017tfx}.
\textit{Anomaly detection} techniques, such as Isolation Forests~\cite{liu2008isolation} and One-Class SVM~\cite{scholkopf2001estimating}, are used to identify data points that deviate significantly from the norm, indicating potential errors or outliers in the data. More recent methods, like Deep Anomaly Detection using Autoencoders and LSTM-based models~\cite{chauhan2015anomaly}, provide enhanced detection capabilities for complex, high-dimensional data.

Automated validation tools are increasingly used to maintain data quality in large-scale ML systems. Schelter et al.~\cite{schelter2018automating} introduced Deequ, a validation tool for extensive datasets using Apache Spark, which allows user-defined quality constraints and provides an API for real-time monitoring through incremental assessments (cf. Section 4.4). Deequ includes ML techniques for automating tasks like outlier detection and value prediction.
Data Linter~\cite{hynes2017data} provides automated suggestions for data cleaning by analyzing data schemas, statistics, distributions, and data type inconsistencies that could impact model training.
Google’s TensorFlow Extended (TFX)~\cite{MLSYS2019_5878a7ab}, deployed in Google Play’s recommendation systems, includes schema inspection and anomaly detection tools to ensure data quality throughout the training and deployment phases.

\textit{Data cleaning} techniques address other challenges such as noise, outliers, and duplicate records during validation. While
there is a large literature on data cleaning~\cite{rekatsinas2017holoclean,li2021cleanml,krishnan2016activeclean},  
not all the cleaning techniques directly benefit ML accuracy. Cleaning techniques are more effective when they target improving model accuracy and making training more robust to noise. Indeed, data noise is considered adversarial when it contains malicious poisoning. In such cases, cleaning involves data sanitization techniques aimed at removing malicious inputs, along with outlier detection methods~\cite{boukerche2020outlier} like Isolation Forests and deduplication~\cite{xia2016comprehensive,meyer2012study} processes to eliminate anomalies and duplicate records, thereby further enhancing the robustness of machine learning models.

\textit{Data imputation} fills gaps caused by missing data, thereby enhancing the completeness and consistency of training data. 
Techniques such as Multiple Imputation by Chained Equations (MICE)\cite{ troyanskaya2001missing} perform multiple imputations to generate several complete datasets, while k-NN~\cite{10.1016/j.jss.2012.05.073} iteratively imputes values by calculating distances for both numerical and categorical attributes. Generative adversarial networks (GANs)~\cite{pmlr-v80-yoon18a}, adapted for imputation, generate synthetic values that accurately fill data gaps, while autoencoder-based models~\cite{pmlr-v97-mattei19a} leverage latent variable techniques to impute data missing at random. Additionally, recent adaptive frameworks automatically select optimal models and hyperparameters, further improving imputation quality~\cite{conf/icml/JarrettCLCS22}.

\subsubsection{Data Augmentation.} Data augmentation techniques allow for improving the robustness of models, particularly for datasets that are small or imbalanced. They expand training data to increase their diversity, mitigate overfitting, and maintain model performance. 
For imbalanced datasets, techniques such as SMOTE~\cite{chawla2002smote} and Mixup~\cite{chawla2002smote} are commonly used. SMOTE works by creating synthetic samples for underrepresented classes, while Mixup involves mixing pairs of examples from different classes to generate new training samples.
Deep learning-based techniques like Variational Autoencoders (VAEs) ~\cite{kingma2013auto} and GAN-based data augmentation~\cite{goodfellow2014generative} models such as StyleGAN~\cite{karras2019style} have further advanced the quality and diversity of synthetic data. However, GANs are limited in generating data that is drastically different from the original. Methods like AutoAugment~\cite{chawla2002smote} attempt to mitigate this by automating augmentation strategies and improving the balance between realistic and diverse data.
Finally, quality assessment metrics such as signal-to-noise ratio and data distribution comparisons are applied to ensure that the augmented data maintains a balance between authenticity and diversity.


\subsubsection{Data Labeling. } Supervised learning relies heavily on high-quality labeled data. 
Several advanced techniques and tools have been developed to \textit{automate data labeling and validation}, significantly enhancing efficiency while ensuring consistency and accuracy. 
For instance, Snorkel~\cite{ratner2017snorkel} uses weak supervision to automatically label large datasets, leveraging heuristic rules, domain knowledge, and other resources to create high-quality training data.  
Prodigy~\cite{prodigy}, a tool developed by Explosion AI, uses active learning to optimize the labeling process, presenting the most informative samples to human annotators to improve efficiency and accuracy. 



\subsubsection{Fairness and Bias Metrics Checking.} Addressing fairness involves both the detection and mitigation of biases in datasets, algorithms, and model outputs. Various fairness metrics are used to assess the model's predictions across several demographic groups, and throughout different stages of ML pipelines (before/in/post processing)~\cite{neutatz2021cleaning}. Fairness metrics include more often individual fairness, where everyone is treated similarly; or group fairness where differentiation is within groups (e.g. women vs. men) or between groups (e.g. young women vs. black men)~\cite{dwork2012fairness}. The most common metrics are: disparate impact~\cite{feldman2015certifying}, equal opportunity~\cite{hardt2016equality}, and demographic parity~\cite{dwork2012fairness}. 
Bias mitigation techniques, such as reweighting samples~\cite{kamiran2012data} during training and adversarial debiasing~\cite{zhang2018mitigating}, have been developed to actively adjust the model's learning process. Reweighting modifies sample weights to compensate for imbalances, while adversarial debiasing introduces an adversarial component to train the model to make fairer predictions. These techniques ensure that fairness is preserved throughout the development stage of ML pipelines. Tools such as IBM’s Fairness 360~\cite{aif360} and Google’s What-If Tool~\cite{what_if_tool} provide frameworks to test and mitigate bias in machine learning models, allowing practitioners to monitor fairness issues that may arise from data drift or changes in societal norms.

\subsubsection{ML Monitoring.}
Once a model is deployed, the focus on serving data should continue. At this stage, the focus of quality improvement shifts to monitoring the properties of serving data and ensuring that they are contextually similar to the training data. Various techniques target the issue of training-serving skew. ~\cite{polyzotis2018data, schelter2018automating} proposed analyses to detect this issue in pre-defined variables. However, determining which variables to monitor and setting the appropriate thresholds for monitoring remains an open challenge. One solution is to base these thresholds on the expected distribution of relevant features (e.g., the usage frequency of specific attributes or demographics), allowing for more targeted monitoring.
Automated monitoring systems can provide additional support by issuing alerts for anomalies, data integrity issues, or deviations in performance metrics like accuracy, precision, and recall.
These automated processes are particularly effective when combined with performance-tracking tools, such as control charts—a key tool from statistical process control (SPC)~\cite{montgomery2019introduction}—which track and visualize fluctuations in model performance over time. 
In addition, techniques used during data preparation, such as schema validation and anomaly detection, can be adapted for continuous data validation in production environments. This ensures the data fed into the model consistently meets the expected quality standards. 

\subsubsection{Data Drift Detection and Adaptation.} In production environments, data distributions may change over time, a phenomenon known as drift. Detecting and adapting to drifts is crucial for maintaining model accuracy and relevance. Various techniques have been developed to address this issue.
Statistical methods are commonly used for detecting changes in data distribution. The \textit{Kolmogorov-Smirnov Test}~\cite{fasano1987multidimensional} is a non-parametric test used to compare the distributions of two datasets and detect significant changes. 
Machine learning-based methods offer more sophisticated approaches to drift detection. The \textit{Maximum Mean Discrepancy (MMD)}~\cite{gretton2012kernel,chwialkowski2015fast} is a kernel-based statistical test that measures the difference between the distributions of two samples. This method is effective for detecting subtle changes in data distribution. Another method, the \textit{Classifier Two-Sample Test (C2ST)}~\cite{lopez2016revisiting}, involves training a classifier to distinguish between two datasets; significant differences in classifier performance indicate drift. The \textit{MMD-D}~\cite{liu2020learning} extends MMD for detecting drift in a more focused manner, providing enhanced detection capabilities.

Adaptation techniques are essential for responding to detected drifts and maintaining model performance. \textit{Reweighting}~\cite{bickel2009discriminative} adjusts the weights of samples during training to compensate for distribution changes, ensuring that the model remains accurate despite shifts in the data. \textit{Incremental Learning}~\cite{polikar2001learn++} involves continuously updating the model with new data to adapt to changing distributions, making the model robust to evolving data patterns. Additionally, \textit{Adversarial Training}~\cite{ganin2016domain} trains models using adversarial examples to make them more robust to distribution changes, enhancing the model's ability to generalize across different data distributions. Another important technique is \textit{Active Learning}~\cite{settles2009active}, which involves selectively querying the most informative data points for labeling and retraining, thereby efficiently updating the model to adapt to new data distributions and reducing the labeling effort required.

\begin{figure}[tb]
   \centering
    \includegraphics[width=1\columnwidth]{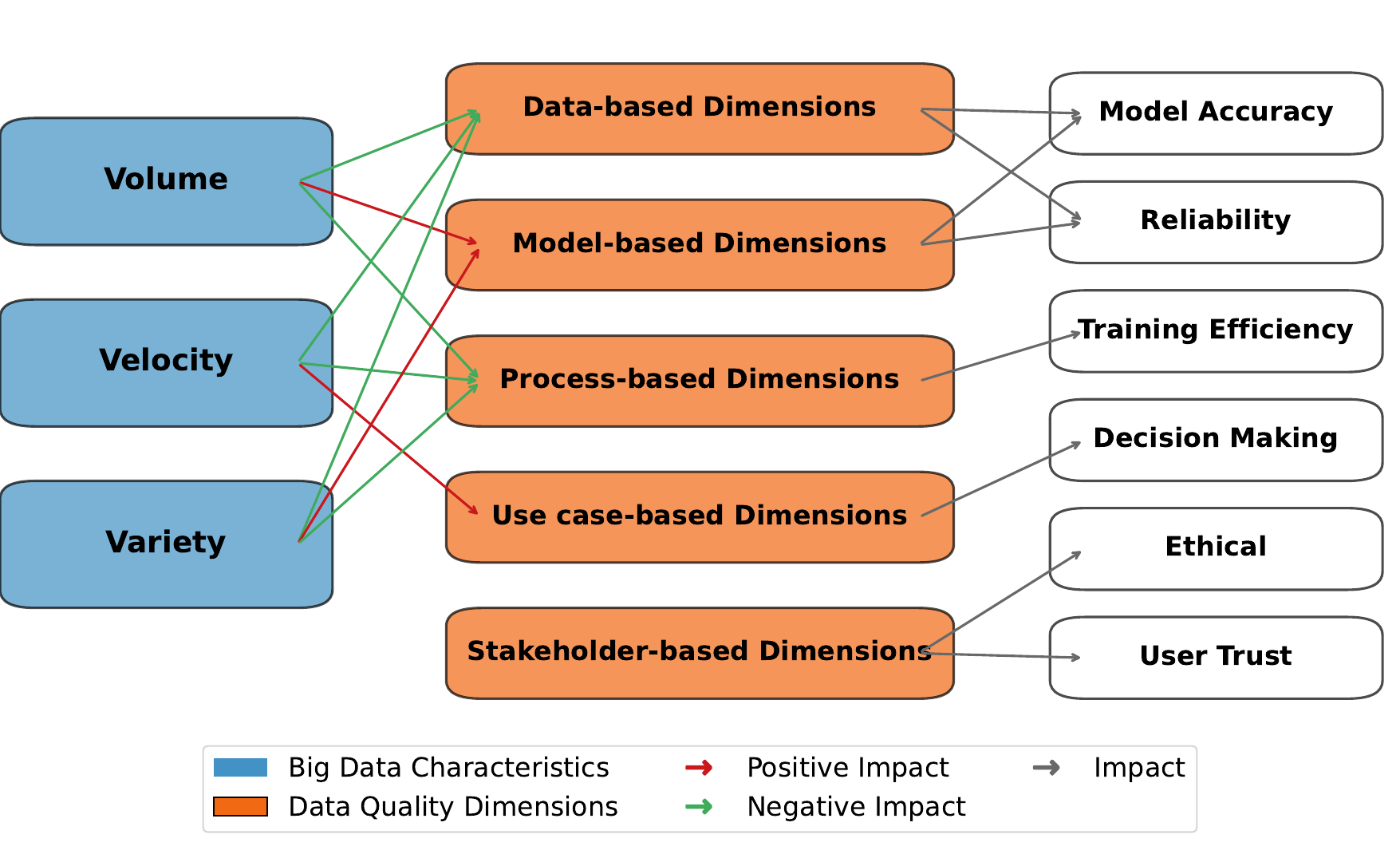}
    \vspace{-1.2\baselineskip}
   \caption{DQ impact in ML Pipelines}
   \label{fig:impact}
\end{figure}

\subsection{Impact of Data Quality in ML Pipelines}
The ML-based data quality dimensions may impact the different stages of ML pipelines which influence the intermediate processes and decision-making outputs. 
They affect how the data is prepared and modeled, which in turn shapes the quality of predictions and decisions made by ML models. This effect extends beyond technical performance to affect decision-making processes. 
Decisions driven by poorly trained models can lead to significant effects, including financial losses, reputational damage, and even risks to safety in critical applications.
Another crucial aspect to consider when assessing the quality in ML pipelines is the interconnected nature of quality dimensions
\cite{d2022modeling,zhou2024surveydataqualitydimensions, 10.1145/3592616}. For instance, accuracy, reliability, and completeness influence each other across different stages of the ML pipeline.
One of the most direct impacts of poor data quality is on model \textit{accuracy}. 
When models are trained on incorrect or biased data, they tend to replicate these inaccuracies in their predictions. This propagation of error can lead to decisions that are not only wrong but potentially harmful, especially in high-sensitive domains like healthcare or finance.

\textit{Reliability} in ML models depends significantly on the consistency and completeness of training data. Inconsistencies, such as duplicates, can lead to unpredictable model behavior, while incomplete data can cause the models to miss critical underlying patterns, significantly reducing their reliability and general applicability. This will affect the \textit{efficiency}, as poor data quality may extend training time and increase computational costs; and also \textit{compliance} leading to legal and ethical risks. Lastly, user trust and adoption of ML technologies by users are closely related to the perceived reliability and accuracy of these systems. Poor data quality can erode user trust, resulting in lower adoption rates and limited impact of these technologies.

The big data challenges amplify the effect of poor data quality in ML pipelines \cite{7906512,10.14778/3415478.3415562}. In ML pipelines, the big data challenges introduce both challenges and opportunities for maintaining data quality throughout the pipeline's stages. These challenges influence key ML-based DQ dimensions like correctness, completeness, timeliness, and consistency, all essential for generating trustworthy decision-making outputs.
In ML pipelines, similar quality awareness techniques, such as sampling techniques and incremental profiling, are applied to manage large volumes of training data. These techniques help address issues like data redundancy and noise, ensuring that models are trained on high-quality data while filtering unnecessary low-value information.
In addition, variety and veracity awareness address trustworthiness issues generated from inconsistent or unreliable big data sources and can be directly applied to ML pipelines by validating training data against established trust metrics. These techniques help maintain model reliability and prevent biased decision-making due to poor data veracity. 
Similarly, the challenge of managing data drift in ML pipelines can be addressed by adapting techniques like real-time anomaly monitoring, which are commonly used in big data systems. By implementing these monitoring practices for serving data in deployed ML models, systems can detect and respond to data drift, ensuring models are retrained as needed to maintain relevance and accuracy over time. Figure \ref{fig:impact} depicts the data quality impact in ML pipelines, considering the big data and ML quality challenges.

Despite these overlapping techniques, there remains a gap between the quality awareness techniques for big data and the quality challenges faced in ML pipelines. For instance, while big data systems are adept at handling massive datasets, they may lack the sophisticated fairness and explainability tools needed to ensure ethical ML model behavior. Similarly, techniques that work well for ensuring data veracity in big data environments might not fully address the complexities of bias detection and mitigation required for maintaining fairness in ML decision-making processes. 

In addition, cleaning and validation methods from big data can unintentionally harm fairness in ML decision-making. As presented in \cite{f077760995674a579d8289d57dc46aa2}, automated data cleaning techniques may remove outliers or noise, potentially excluding critical data points that represent marginalized groups, leading to less representative and biased models. Similarly, validation processes based on historical data may reinforce existing biases and societal inequalities. This underscores the necessity for tailored quality awareness techniques that address both the challenges of big data and prioritize fairness in ML systems. Closing this gap is vital to ensure that data quality methods do not compromise the integrity of ML models and their decisions.

The cause-effect connection between big data and ML is evident in the way big data quality challenges are amplified within ML pipelines. By adopting and adapting quality awareness techniques from big data, such as sampling, profiling, and quality checks, it is possible to mitigate some of these challenges in ML pipelines. However, the growing demands for fairness, explainability, and ethical alignment in ML also necessitate the development of more tailored techniques that go beyond the traditional techniques used in big data systems.

%% file: Sections/OpenChallenges.tex
\section{New Opportunities of Data Quality Awareness for Data Science Systems}

The intersection of big data and machine learning within data science systems presents unique challenges in data quality, which opens up several opportunities for advancing quality awareness techniques. This section explores potential research directions arising from these challenges, particularly focusing on the impact of large language models (LLMs) and the integration of ethical considerations into data science practices.

\subsection{Enhancing Quality Awareness in DS Systems}

\textit{Challenges with Multimodal Data:}
Current data quality systems struggle with the diverse characteristics of multimodal data, such as discrepancies between discrete text and continuous image data~\cite{li2017multi}. Synchronization issues in time-dependent data formats like video and audio further complicate accurate model performance. Future research should develop unified quality metrics and cross-modal consistency checks tailored for multimodal environments~\cite{ye2022noise}. Advanced data fusion techniques, such as multi-modal GANs, could address data gaps and reduce redundancy, ensuring robust data integrity across different modalities~\cite{zhang2021partial}.

\textit{Adaptive Quality Monitoring:}
Dynamic environments where data continuously evolves, such as those involving real-time data streams, demand adaptive quality monitoring frameworks. Integrating ML models with dynamic profiling techniques could facilitate real-time adaptation to emerging data patterns like drift, enhancing the responsiveness of DQ validation systems~\cite{schelter2018automating, MLSYS2019_5878a7ab}.

\subsection{Addressing AI Ethics}
An important research direction within the intersection of fairness and data cleaning is addressing how existing quality awareness techniques affect fairness in ML pipelines. Specifically, we need to examine whether existing data cleaning techniques can be effectively adapted to handle fairness constraints, or if new techniques should be developed. One promising opportunity lies in extending current cross-validation and model optimization techniques to include fairness, rather than only prioritizing accuracy. Techniques like Shapley values~\cite{karlavs2024data} have been proposed to identify data points that negatively impact fairness, suggesting a path for fairness-enhanced cleaning. The big data challenges further amplify fairness issues in ML pipelines, emphasizing the need for scalable, fairness-aware solutions that balance accuracy and ethical considerations. In addition, enhancing transparency and interpretability in fairness-aware cleaning is also a key research direction to leverage trust and accountability in data science systems~\cite{holzinger2019causability, wu2024usable}. Developing techniques that clarify how these fairness-aware cleaning impact model fairness will help reveal features affecting outcomes.

Furthermore, collaborative efforts that include legal and social science perspectives are essential to developing fairness metrics that reflect diverse social, legal, and ethical contexts~\cite{arrieta2020explainable, cheng2021socially}. Interdisciplinary approaches can enrich technical solutions with broader societal insights, helping to refine fairness definitions and implement more effective mitigation strategies.


\subsection{Quality Impact of Large Language Models}

LLMs have become integral to data science systems, offering the ability to extract insights and make data-driven decisions at scale. However, one of the main challenges associated with LLMs is the issue of "hallucination"~\cite{ji2023survey,bender2021dangers}
This issue often arises from harmful errors in the original training data, where biases or inaccuracies lead to hallucinations, and can be exacerbated when LLMs are trained on outputs generated by other LLMs, which creates a feedback loop that further degrades output quality~\cite{tonmoy2024comprehensive}.

Addressing these challenges requires ensuring the quality and authenticity of the data used in both training and evaluation~\cite{christiano2017deep, zhao2024expel}. While conventional methods such as verifying data sources, using human-in-the-loop validation and ensuring transparency in data provenance, help mitigate hallucination risks. However, the scale and complexity of LLMs require new approaches to effectively manage and improve data quality as these models continue to expand~\cite{lewis2020retrieval, ji2023survey}. Additionally, new data quality metrics may be required to capture the specific challenges introduced by LLMs, including mechanisms to verify generated content and manage internal biases, ensuring accuracy and reliability as they scale.

%% file: Sections/conclusion.tex
\section{Conclusions}
\label{sec:concl}
This paper reviews the evolution of data quality awareness, examining its progression from traditional data management to contemporary data science, focusing on big data and machine learning contexts. 
We identified key challenges in big data, such as managing extensive volumes, rapid data influx, and varied types, necessitating adaptive and scalable quality assessment. 
Additionally, we explored the impact of data quality within ML pipelines, demonstrating its influence on model accuracy and reliability, and discussed techniques like real-time validation, drift detection, and data augmentation to maintain model integrity. 
We also explored new opportunities in data quality for data science.